\documentclass[11pt,hidelinks,colorlinks=true,linkcolor=blue,citecolor=blue,urlcolor=blue]{iopart}
\usepackage{iopams,mathrsfs,color}

\expandafter\let\csname equation*\endcsname\relax
\expandafter\let\csname endequation*\endcsname\relax

\usepackage{orcidlink}
\usepackage[utf8]{inputenc}
\usepackage{graphicx}  
\usepackage{bm}
\usepackage{soul}
\usepackage{dsfont}
\usepackage{enumitem}
\usepackage{physics}
\usepackage{amsfonts,amsmath,amssymb,mathtools} 
\usepackage{hyperref}
\usepackage{orcidlink}

\newcommand{\ddd}{\mathrm{d}}

\begin{document}
\title{A spectral viewpoint on the single defect tight-binding chain}
\author{Sayan Roy
\orcidlink{0009-0005-5805-3791}}
\address{Theoretical Physics,  Saarland University,  D-66123  Saarbr\"ucken,  Germany}
\address{Center for Quantum Technologies (QuTe), Saarland University, Campus, 66123 Saarbr\"ucken, Germany}


\begin{abstract}
    We analyze the time evolution of the nearest-neighbour tight-binding chain in the presence of a single onsite defect. Such a defect was shown to generate non-trivial transport behavior recently in the article by \textit{Acharya et al J. Stat. Mech. (2026) 043102}. The authors have used a defect technique inspired by classical random walk methods to obtain exact analytical expressions for the occupation probability and subsequently the mean and mean-squared displacement (MSD). Here we derive the same results using a spectral decomposition approach. Starting from the secular equation, we obtain the self-consistency condition for the eigenvalues and construct the corresponding normalized eigenvectors. This approach naturally separates the Hilbert space into a dark subspace, whose states have zero amplitude at the defect site and remain unaffected, and bright subspaces, whose states get modified because of the defect. Using this eigenvalue decomposition, we provide a spectral origin of non-monotonicity in the MSD. Numerical calculations for finite chains show that the analytically estimated critical defect strength is in excellent agreement with the defect strength that minimizes the MSD.
\end{abstract}
\maketitle

\section{Introduction}
Understanding transport properties in disordered systems is one of the central problems in current quantum technologies. The paradigmatic theoretical model to describe disordered systems is the celebrated Anderson model, in which uncorrelated onsite disorder suppresses wave propagation through destructive interference~\cite{Anderson1958}. However, significant changes in quantum transport do not necessarily require disorder throughout the system. Even a single local impurity, commonly called a defect, may lead to decoherence in different platforms and alter the propagation of wavefunctions~\cite{Romaneko17, Gregory_2002, Weeden2026}. 

The analytical Green-function treatment of such a localized perturbation dates back to the Koster--Slater method~\cite{Koster1954} and Lippmann-Schwinger/T-matrix treatments~\cite{Lippman_1950}, in which impurity-induced eigenvalues are determined from the diagonal resolvent of the defect-free lattice. Recently, Acharya \textit{et al.} studied this problem for a periodic tight-binding chain containing a single onsite defect and involving a particle hopping between the nearest-neighbour sites of a lattice~\cite{AcharyaGiuggioliGupta}. Using a defect method adapted from classical random-walk theory, they obtained an exact time-dependent expression for the occupation probability of the particle at different lattice sites. At the algebraic level, their formalism is closely related to the Koster--Slater construction~\cite{Koster1954}, since both reduce the single-defect problem to a Green-function resolvent identity at the defect site. The results of Ref.~\cite{AcharyaGiuggioliGupta} show that the effect of the defect depends strongly on the initial position of the particle. When the particle starts at the defect site, its spreading is suppressed monotonically as the defect strength increases. In contrast, when the particle starts away from the defect, the mean-squared displacement (MSD) depends non-monotonically on the defect strength. 

The single-defect model can also be viewed from the complementary perspective of a continuous-time quantum-walk search problem. In this interpretation, the defect site plays the role of a marked state, and the objective is to transfer probability from an initially uniform superposition state to the marked state. The search is governed by the spectral gap which determines the characteristic timescale of the process~\cite{ChildsGoldstone}. Motivated by this connection, we employ the secular equation and eigenstate decomposition approach used in quantum spatial search problems to provide a complementary derivation of the results obtained in Ref.~\cite{AcharyaGiuggioliGupta} and additional physical insight into their origin. The formulation relies only on the principles of quantum mechanics without the need to invoke the defect technique, and makes the underlying spectral structure more transparent. Our central question is how the different spectral sectors determine the long-time MSD as the defect strength varies. We resolve the dynamics into defect-independent dark states and extended and localized bright states, and quantify their contributions to the MSD. This decomposition explains the non-monotonic behavior and shows why the defect strength that maximizes the initial state's overlap with the localized state closely predicts the MSD minimum.

The paper is organized as follows. In Sec.~2, we introduce the model and construct its dark and bright eigenstates. In Sec.~3, we use this eigenbasis to derive the occupation probability and connect our formulation to Ref.~\cite{AcharyaGiuggioliGupta}. In Sec.~4, we examine the contributions of the spectral sectors to the long-time MSD, explain its non-monotonic dependence on defect strength, and assess the finite-size effects. Section~5 summarizes our conclusions. Further technical details and an extension to multiple defects are presented in the appendices.

\section{Spectral decomposition of the single defect tight-binding chain}
We consider a one-dimensional tight-binding chain of $N$ sites, labeled by $n=0,1,\ldots,N-1$, with a single onsite defect at site $n_d$. The dynamics is governed by the Hamiltonian
\begin{align}
    \hat{H} = \hat{H}_0-q\ket{n_d}\bra{n_d},
    \label{Eq:H_tot}
\end{align}
where $q> 0$ denotes the defect strength and
\begin{align}
    \hat{H}_0 = -\gamma\sum_{n=0}^{N-1} \left( \ket{n+1}\bra{n} + \text{H.c.}\right)
    \label{H_0}
\end{align}
is the defect-free Hamiltonian. We impose periodic boundary conditions such that $\ket{N}\equiv\ket{0}$, and $\gamma>0$ is the hopping amplitude. Throughout the manuscript, we set $\hbar=1$. The particle is initially localized at site $n_0$, such that $\ket{\psi(0)}=\ket{n_0}$. We define the distance between any two sites on the ring as $d(n,m)\equiv\min \{|n-m|,N-|n-m|\}$ and denote the distance between the initial and defect sites by $l=d(n_0,n_d)$.

The eigenstates of the defect-free Hamiltonian $\hat H_0$ are the Bloch states~\cite{ashcroft_solid_1976}
\begin{align}
  \ket{k_r} =  \frac{1}{\sqrt{N}} \sum_{n=0}^{N-1}  \e^{i k _r n}\ket{n},
\end{align}
where $ k_r = \frac{2\pi r}{N}$, and $r=0,1,\dots,N-1$. The corresponding eigenenergies are
\begin{align}
    \epsilon_{k_r} =-2\gamma \cos k_r.
\end{align}
Except for the state with $r=0$ and, when $N$ is even, the state with $r=N/2$, the eigenenergies are twofold degenerate and occur in pairs $\{k_r,k_{N-r}\}$.

The central quantity of interest is the mean-squared displacement (MSD) about the initial site which is defined as
\begin{align}
\Delta_2^{(d)}(t)=\sum_{n=0}^{N-1}d(n,n_0)^2\,P_n^{(d)}(t).\label{eq:MSD_time}
\end{align}
where  $P_n^{(d)}(t)$ denotes the occupation probability at site $n$ and time $t$, which can be computed by
\begin{align}
P_n^{(d)}(t)=\left\vert\bra n \e^{-i\hat Ht}\ket{n_0}\right\vert^2.
\label{eq:occupation_probability}
\end{align}
For $q>0$, the defect lifts the relevant degeneracies of the defect-free spectrum (see also Fig.~\ref{fig:Fig1}(a)). Denoting the complete orthonormal set of eigenstates of $\hat H$ by ${\ket{\phi_j}}$, the infinite-time-averaged occupation probability and MSD therefore take the diagonal-ensemble form
\begin{align}
    \overline P_n^{(d)} &=\lim_{T\rightarrow\infty}
\frac{1}{T}\int_0^T \mathrm{d}t\, P_n^{(d)}(t)=  \sum_{j} \left\vert\langle n\vert\phi_j\rangle\right\vert^2\, \vert \langle \phi_j \ket{n_0}\vert^2 ,  \label{eq:Pbar_eigenstates} \\
\overline \Delta_2^{(d)} &= \lim_{T\rightarrow\infty}
\frac{1}{T}\int_0^T \mathrm{d}t\, \Delta_2^{(d)}(t)=\sum_{n=0}^{N-1}d(n,n_0)^2\,\overline P_n^{(d)}
\label{eq:MSD_main}
\end{align}
respectively. Throughout the analysis for $q>0$, the limit $T\rightarrow\infty$ is taken at fixed nonzero $q$, whereas the point $q=0$ should be treated separately since at $q=0$, terms within each degenerate eigenspace do not dephase under the time average~\cite{AcharyaGiuggioliGupta}. Since the states become degenerate as $q\rightarrow0^+$, the limits $T\rightarrow\infty$ and $q\rightarrow0$ need not commute.

\subsection{Self-consistency equation for eigenvalues}
In this section, we show how the spectrum of $\hat H$ can be understood in terms of the spectrum of $\hat H_0$ using a self-consistency approach closely connected to the localized-impurity treatment in Ref.~\cite{Koster1954}. This approach has also been used to study analog Grover search problems in different graph topologies~\cite{ChildsGoldstone,Emma2025}, non-Hermitian search protocols~\cite{king2026timecomplexitymonitoredquantum} and, more recently, to study relaxation dynamics in a tight-binding chain in the presence of dephasing and nonlinear defects~\cite{das2026dephasinginducedrelaxationtightbindingchains}. Here, we use the technique to uncover the different classes of eigenstates and the roles they play in determining the MSD. Let $\ket{E}$ be an eigenstate of $\hat H$ with a nonzero defect amplitude, $u_{n_d}(E) = \braket{n_d}{E}\neq 0$. Such eigenstates couple to the defect and will be referred to as bright states. By contrast, eigenstates of $\hat{H}$ with zero amplitude at the defect site are obviously unaffected by the presence of the defect and form the dark state subspace. Consider now the eigenvalue equation for $\hat H$ in the bright subspace:
\begin{align}
  (\hat H_0-q\ket{n_d}\bra{n_d})\ket{E}=E\ket{E} \implies
  (\hat H_0- E)\ket{E} =  q\, u_{n_d}(E) \,\ket{n_d} \,.
  \label{eq:eigvec}
\end{align}
Inverting the above equation~\eqref{eq:eigvec} gives 
\begin{align}
    \ket{E}&= \frac{q\, u_{n_d}(E)}{\hat H_0 - E} \ket{n_d}\,.
    \label{eq:eigvec_nd}
\end{align}
Since we consider $u_{n_d}(E) \neq 0$, projecting Eq.~\eqref{eq:eigvec_nd} onto $\bra{n_d}$, one obtains the \textit{self-consistency condition} for a bright eigenvalue as $q F(E) = 1$, where we have defined $F(E) \equiv \bra{n_d} (\hat H_0 - E)^{-1}  \ket{n_d}$. This self-consistency condition is a generic consequence of the rank-one form of the perturbation and is valid for any arbitrary unperturbed Hamiltonian $\hat H_0$; the lattice geometry and hopping structure enter only through $F(E)$~\cite{Koster1954, ChildsGoldstone, Emma2025}. Inserting $\sum_r \ket{k_r}\bra{k_r} = \mathds{1}$ and using the defect-free eigenenergies $\epsilon_{k_r}$, the self-consistency condition becomes
\begin{align}
  1- qF(E) =   1 + \frac{q}{N} \sum_{r=0}^{N-1} \frac{1}{E+2\gamma\cos(2\pi r/N)} &=  0.
\label{eq:secular_eqn_tbm}
\end{align}
\subsection{Bright and Dark eigenstates}
\label{sec:bright}
We construct the normalized bright and dark eigenstates in this section. The normalization condition $\bra{E}\ket{E} = 1$ for a bright eigenstate $\ket{E}$ in Eq.~\eqref{eq:eigvec_nd}  gives $ |u_{n_d}(E) |^2\bra{n_d}\,q^2\,(\hat H_0 - E)^{-2}\ket{n_d} = 1$, which can be rewritten as 
\begin{align}
    |u_{n_d}(E)|^2 =  \frac{1}{q^2 F^{\prime}(E)},
    \label{eq:u_{n_d}}
\end{align}
where $F^{\prime}(E)$ is the derivative. Substituting Eq.~\eqref{eq:u_{n_d}} into Eq.~\eqref{eq:eigvec_nd}, we get the components of normalized bright states as
\begin{align}
u_n(E) =\frac{\langle n|(\hat H_0-E)^{-1}|n_d\rangle}{\sqrt{F'(E)}} = \frac{1}{\sqrt{F'(E)}}\,\frac{1}{N}\sum_{r=0}^{N-1}
\frac{e^{ik_r(n-n_d)}}{\epsilon_{k_r}-E}.
\label{eq:bright}
\end{align}
The bright states can be further classified according to their energies and spatial profiles. The bright eigenstates whose energies lie inside the defect-free energy band, $-2\gamma\leq E_j\leq 2\gamma$, remain spatially extended over the lattice and will be referred to as\textit{ bright extended states}. In contrast, the self-consistency equation yields a bright eigenvalue that lies below the band, i.e., $E_{j}<-2\gamma$; the corresponding eigenstate is localized around the defect and will be referred to as the \textit{bright localized state}. The uniqueness of the outside of the band eigenvalue follows more generally from the rank-one structure of the perturbation. The Cauchy interlacing theorem implies that at most one eigenvalue can occur below the lower edge of the unperturbed spectrum, and no eigenvalue appears above the upper edge of the unperturbed spectrum~\cite{Hwang2004CauchysIT}. These conclusions are generic and applies to any arbitrary graph topologies.

Eigenstates of $\hat{H_0}$ that are orthogonal to $|n_d\rangle$ are unaffected by the defect term. For each degenerate pair $\{k_r,k_{N-r}\}$, one may construct a normalized \textit{dark state} $\ket{D_r}$, whose components are given by~\cite{thiel20, roy25}
\begin{align}
v_n(\epsilon_{k_r}) = \bra n D_r \rangle =\sqrt{\frac{2}{N}}\,\sin \big[k_r(n-n_d)\big]\,,
\label{eq:dark}
\end{align}
and the corresponding eigenvalue equation gives $\hat H|D_r\rangle= \hat H_0 |D_r\rangle =\epsilon_{k_r}|D_r\rangle$. Thus, each degenerate pair gives one dark state giving a total of $\lfloor(N-1)/2 \rfloor$ dark states. It also implies that $\ket{D_r}$ is a common eigenstate of $H$ and $\hat H_0$. This can also be understood from the perspective of reflection symmetry about the defect site. Define the reflection operator $\hat{R}$ by its action $\hat{R} \ket{n_d + r} = \ket{n_d  - r}$. It then follows directly from Eq.~\eqref{eq:dark} that $\hat{R}\ket{D_r}=-\ket{D_r}$, showing that the dark states are odd under reflection about the defect site and therefore have zero amplitude at the defect site. Consequently, the perturbation term $-q\ket{n_d}\bra{n_d}$ does not affect these states. Thus, reflection symmetry explains the specific form of the dark states in the ring, whereas, more generally, dark states are eigenstates $\ket{\psi}$ of $\hat H_0$  that lie in the kernel of the perturbation operator $\hat V$, satisfying $\hat V\ket{\psi}=0$.

Note that, unlike dark states, the bright states are eigenstate of $\hat H$ only. The poles of the self-consistency equation occur at the defect-free energies $\epsilon_{k_r}$. Hence, the dark eigenvalues shown in Fig.~\ref{fig:Fig1}(a) coincide with a subset of these poles. The eigenstates obtained from \eqref{eq:bright} and \eqref{eq:dark} thus form a complete orthonormal basis. Representative probability distributions for the three classes of eigenstates are shown in Fig.~\ref{fig:Fig1}(b). The time averaged occupation probability can then be written in terms of eigenstates as
\begin{align}
\overline{P}_n^{(d)} &= \sum_{j \in \mathcal{H}_D} \vert \langle n \vert D_j \rangle \vert^2  \, \vert \langle D_j \vert n_0 \rangle \vert^2 + \sum_{j\in \mathcal{H}_B^{\rm ext}}  \vert \langle n \vert E_j \rangle \vert^2  \, \vert \langle E_j \vert n_0 \rangle \vert^2 +  \vert \langle n \vert E_{\rm loc} \rangle \vert^2  \, \vert \langle E_{\rm loc} \vert n_0 \rangle \vert^2, \nonumber \\
&= \overline{P}_{n,\textrm{dark}}^{(d)} + \overline{P}_{n, \textrm{extd.}}^{(d)} + \overline{P}_{n, \textrm{loc.}}^{(d)}
\label{eq:MSD_spectral}
\end{align}
where $\mathcal{H}_D$ and $\mathcal{H}_B^{\rm ext}$ denote the dark and bright extended subspaces, respectively. The first term is the contribution of the dark states, the second term is the contribution from the bright extended states and the third term is the contribution from the bright localized state. Consequently, the infinite time-averaged MSD can be written as $\overline{\Delta}_2^{(d)} = \overline{\Delta}_{2,\textrm{dark}}^{(d)} + \overline{\Delta}_{2, \textrm{extd.}}^{(d)} + \overline{\Delta}_{2, \textrm{loc.}}^{(d)}$. We will discuss the contribution of each term in more detail in Sec.~\ref{sec:localized_bright_state}.
\subsection{Recasting the self-consistency equation in terms of Chebyshev polynomials}
\begin{figure}[!htpb]
\centering
\includegraphics[width = \linewidth]{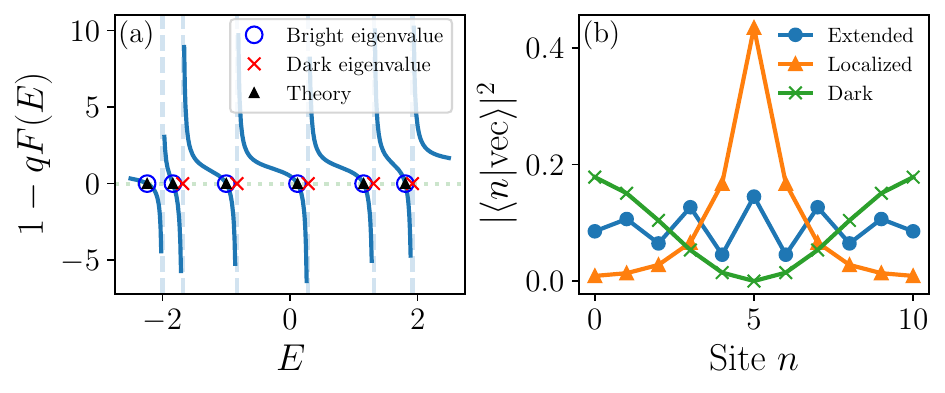}
\caption{(a) $1 - qF(E)$ is plotted as a function of $E$. The defect-free eigenvalues form the poles of the equation, which are shown by the vertical dashed lines. Except for the ground state eigenvalue, half of the degenerate spectrum forms the dark subspace. The bright eigenvalues are formed when the self-consistency equation $1- qF(E) = 0$ is satisfied. The blue circles (bright eigenvalues) and the red crosses (dark eigenvalues) are obtained by exact diagonalization of the Hamiltonian. The bright eigenvalues match  the black triangles, which are the roots of the polynomial equation $Q_{B}^{\rm odd}(x)$ (See \ref{app:1}). (b) Occupation probability as a function of the site index for three different types of states. Other numerical parameters are $N = 11$, $\gamma = 1$, $q = 1$, $\ket{n_d} = \ket{5}$. }
\label{fig:Fig1}
\end{figure}

To obtain an explicit finite-size polynomial form of the self-consistency equation, we rewrite Eq.~\eqref{eq:secular_eqn_tbm} in terms of Chebyshev polynomials. This representation also allows us to establish its correspondence with the pole condition obtained in Ref.~\cite{AcharyaGiuggioliGupta}. We denote the Chebyshev polynomial of the first and second kinds by $T_N(x)$ and $U_N(x)$, respectively. Defining $x = - E/2 \gamma$, we write the secular equation ~\eqref{eq:secular_eqn_tbm} as follows
\begin{align}
    1 - \frac{q}{2 \gamma N}  \sum_{r=0}^{N-1} \frac{1}{x - \cos(2\pi r/N) } = 0\,.
    \label{eq:secular_simplified}
\end{align}
Using the identity $T_N\left(\cos\frac{2\pi r}{N}\right)  =  \cos(2\pi r) =  1$, we see that the numbers $\cos(2\pi r/N)$ are precisely the zeros of $T_N(x)-1$. In other words $T_N(x) - 1 = \Pi_{r= 0}^{N-1} (x - \cos(2\pi r/N))$ up to a constant prefactor. Taking the logarithmic derivative therefore gives
\begin{align}
\label{eq:log_derivative}
  \frac{\ddd}{\ddd x}\log\!\left[T_N(x)-1\right] = \sum_{r=0}^{N-1}  \frac{1}{x-\cos(2\pi r/N)},
\end{align}
which is the sum appearing in the second term in Eq.~\eqref{eq:secular_simplified}. Using Eq.~\eqref{eq:log_derivative}, $T_N'(x)=N U_{N-1}(x)$ and the Pell's identity $T_N^2(x)  -  (x^2-1)U^2_{N-1}(x) = 1$~\cite{NIST:DLMF}, we rewrite Eq.~\eqref{eq:secular_simplified} as
\begin{align}
 1 - \frac{q}{2\gamma}  \frac{U_{N-1}(x)}{T_N(x)-1} =  1 - \frac{q}{2 \gamma} \frac{T_{N}(x) + 1}{(x^2 - 1)U_{N-1}(x)} = 0\,.
   \label{eq:secular_chebyshev}
\end{align}
This forms the polynomial equation
\begin{align}
  Q(x) \equiv  (x^2-1)U_{N-1}(x) -\frac{q}{2\gamma}\left[T_N(x)+1\right]  =  0\,,
  \label{eq:eigval}
\end{align}
which is also the polynomial equation that appears in the denominator of Laplace-domain expression of Ref.~\cite{AcharyaGiuggioliGupta}. However, note that this is a polynomial of degree $N+1$, but the dimension of bright subspace is $N$ minus the dimension of the dark subspace which equals $\lfloor N/2 \rfloor + 1$. Therefore, not all roots of $Q(x)$ can represent bright eigenvalues. In fact, $Q(x)$ contains  $N - \lfloor N/2 \rfloor$ additional roots arising from the common factors of the polynomial $(x^2 - 1)U_{N-1}(x)$ and $T_{N}(x) + 1$. The residues are computed using the roots of the polynomial $Q(x)$ in \cite{AcharyaGiuggioliGupta}, so these additional roots correspond to removable singularities rather than a pole and thus do not contribute to the residue sum. The explicit factorizations for odd and even $N$ are shown in \ref{app:1}. After removing these common factors, the roots of the remaining polynomial equation $x_j$ forms the bright eigenenergies $E_j = -2 \gamma x_j$ of $\hat H$. The self-consistency condition is also checked numerically with exact diagonalization of the total Hamiltonian $\hat{H}$ and is plotted in Fig.~\ref{fig:Fig1}(a).
\section{Occupation probability}
\label{sec:eigvec}
We now derive the occupation probability from the spectral decomposition of $\hat H$. We use the complete bright–dark eigenbasis and show that the defect modifies only the bright sector dynamics. This allows us to define the \textit{defect-induced amplitude} and interpret the different components of the occupation probability obtained in Ref.~\cite{AcharyaGiuggioliGupta}.
\subsection{Transition Amplitude}
We first relate the transition amplitude $\psi^{(d)}(n,n_0,t)$ in the presence of the defect to the corresponding defect-free amplitude $\psi(n,n_0,t)$. We define the projector $\hat \Pi_{ D}=\sum_{j\in\mathcal H_D}|D_j\rangle\langle D_j|$ and its
complement $\hat \Pi_{B}=\mathds{1}-\hat \Pi_{\mathcal D} = \sum_{j\in\mathcal H_B} \ket{E_j}\bra{E_j}$ based on the dark and bright subspaces discussed in Sec.~\ref{sec:bright}. For a particle initially localized at site $n_0$, $\psi(n,n_0,t)$ can be split into contributions from dark and bright sectors as follows:
\begin{align}
    \psi(n,n_0,t) &= \langle n|e^{-i \hat H_0t}|n_0\rangle
    = \langle n|(\hat \Pi_D + \hat\Pi_B)e^{-i\hat H_0t}(\hat\Pi_D + \hat\Pi_B) |n_0\rangle\nonumber \\
     &=  \langle n|\hat\Pi_D e^{-i\hat H_0t}\hat\Pi_D |n_0\rangle +  \langle n|\hat\Pi_Be^{-i\hat H_0t} \hat\Pi_B|n_0\rangle \nonumber \\ &=  \psi_D(n,n_0,t) + \psi_B(n,n_0,t),
\end{align}
where we have used $[\hat\Pi_D,\hat H_0]=0$ and $\hat\Pi_D \hat\Pi_B = 0$ which ensure that the two cross terms vanish. The evolution therefore decomposes into two mutually orthogonal invariant sectors; $\psi_{D}(n,n_0,t)$ and $\psi_{B}(n,n_0,t)$ denote the defect-free transition amplitudes restricted to dark and bright subspaces, respectively. Similarly, in the presence of the defect, the transition amplitude $\psi^{(d)}(n,n_0,t)$ splits into the dark and bright subspaces as $\psi^{(d)}(n,n_0,t)=\langle n|e^{-i\hat Ht}|n_0\rangle =\psi_{D}^{(d)}(n,n_0,t)+\psi_{B}^{(d)}(n,n_0,t)$
where we have used $[\hat\Pi_D,\hat H]=0$, $\psi_{D}^{(d)}(n,n_0,t) = \langle n|\hat \Pi_D e^{-i\hat Ht}\hat\Pi_D |n_0\rangle$ and $\psi_{B}^{(d)}(n,n_0,t) = \langle n|\hat\Pi_Be^{-i\hat Ht}\hat\Pi_B|n_0\rangle$. Since every $|D_j\rangle$ is simultaneously an eigenstate of $\hat H$ and of $\hat H_0$, with the same energy eigenvalue $\epsilon_{k_j}$, the dark sector contribution to the transition amplitude is therefore identical in the presence and absence of the defect. Therefore, $\psi_D^{(d)}(n,n_0,t)=\psi_D(n,n_0,t)$. In other words, the dark subspace is unaffected by the defect and all defect-induced dynamics resides in the bright subspace. To understand its influence, we define the defect-induced amplitude $A(n,n_0,t)$ as the difference between the transition amplitudes with and without the defect:
\begin{align}
A(n,n_0,t) &\equiv \psi^{(d)}(n,n_0,t)-\psi(n,n_0,t) =\psi_B^{(d)}(n,n_0,t)-\psi_B(n,n_0,t),
\label{eq:Adef}
\end{align}
Using this definition, one then obtains the occupation probability in the presence of the defect as $P_n^{(d)}(t)=|\psi^{(d)}(n,n_0,t)|^2$. Writing $\psi^{(d)}(n,n_0,t)=\psi(n,n_0,t)+A(n,n_0,t)$ from \eqref{eq:Adef} and expanding, we obtain,
\begin{align}
P_n^{(d)}(t)= P_n(t) +  I_n^{(d)}(t) + K_n^{(d)}(t),
\label{eq:Pn29}
\end{align}
where we have defined $P_n(t) \equiv |\psi(n,n_0,t)|^2$, $I_n^{(d)}(t)\equiv \psi^{*}(n,n_0,t)A(n,n_0,t)+\psi(n,n_0,t)A^{*}(n,n_0,t)$ and $K_n^{(d)}(t) \equiv |A(n,n_0,t)|^2$.  Equation~\eqref{eq:Pn29} thus resolves the dynamics into three physically distinct contributions. The first term $P_n(t)=|\psi(n,n_0,t)|^2$ is the defect-free propagation. The term $I_n^{(d)}(t)$ describes the interference between the defect-free amplitude and the defect-induced amplitude. It can be either positive or negative, depending on their relative phase, and therefore determines whether the defect enhances or suppresses the probability at a given site. The quantity $K_n^{(d)}(t)$ is the probability carried by the defect-induced component itself. Its spatial structure is controlled by the bright states. The bright extended states spread throughout the lattice, whereas the bright localized state gives a contribution concentrated near the defect. In Sec.~\ref{sec:localized_bright_state}, we quantify how their spectral weights and spatial second moments combine to produce the non-monotonic dependence of the long-time MSD on defect strength.
\subsection{Defect-induced amplitude}
In this subsection, we derive the explicit form of $A(n,n_0,t)$, and show that it is equivalent to the one obtained in Ref.~\cite{AcharyaGiuggioliGupta}. 
We first consider the evolution in the presence of the defect.
Using Eq.~\eqref{eq:u_{n_d}} and Eq.~\eqref{eq:bright}, one finds 
\begin{align}
\psi_B^{(d)}(n,n_0,t) =  \langle n|e^{-i\hat Ht}\hat\Pi_B|n_0\rangle =\frac{q}{N}\sum_{j \in \mathcal{H}_B}\sum_{r=0}^{N-1}\frac{u_{n_d}(E_j)u_{n_0}^{*}(E_j)e^{ik_r(n-n_d)}}{\epsilon_{k_r}-E_j}e^{-iE_jt},
\label{eq:psiBD}
\end{align}
where we have used $[\hat \Pi_B, \hat H] = 0$ and $\hat \Pi_B^2 = \hat\Pi_B$. Although $\ket{E_j}$ are eigenstates of $\hat H$, they are not eigenstates of $\hat H_0$.
Therefore, for computing the defect-free contribution of the same bright component, we first expand $\ket{E_j}$ in the eigenbasis of $\hat H_0$ and  we obtain
\begin{align}
\psi_B(n,n_0,t) = \langle n|e^{-i\hat H_0 t}\hat\Pi_B|n_0\rangle&=\frac{q}{N}\sum_{j \in \mathcal{H}_B}\sum_{r=0}^{N-1}\frac{u_{n_d}(E_j)u_{n_0}^{*}(E_j)e^{ik_r(n-n_d)}}{\epsilon_{k_r}-E_j}e^{-i\epsilon_{k_r} t},
\label{eq:psiB}
\end{align}
where we have used $[\hat \Pi_B, \hat H_0] = 0$. For compactness, let us define
\begin{align}
f_j\equiv u_{n_d}(E_j)u_{n_0}^{*}(E_j).
\label{eq:fj_def}
\end{align}
Using Eqs.~\eqref{eq:psiBD}-\eqref{eq:fj_def}, we write the defect-induced amplitude as
\begin{align}
A(n,n_0,t)&=\frac{q}{N}\sum_{j \in \mathcal{H}_B}\sum_{r=0}^{N-1}\frac{f_j e^{ik_r(n-n_d)}}{\epsilon_{k_r}-E_j}
\left(e^{-iE_jt}-e^{-i\epsilon_{k_r} t}\right).
\label{eq:Aclosed}
\end{align}
The quantity $f_j$ gives information on how much is the overlap of the initial state with the eigenstates of $\hat H$ multiplied by how strong this eigenstate overlaps with the defect site. Thus $qf_j$ is the effective weight of the $j$th bright eigenstate $\ket{E_j}$. In~\ref{app:2}, we have recast $f_j$ in terms of Chebyshev polynomials which exactly matches the expression in Ref.~\cite{AcharyaGiuggioliGupta}. 

\section{Origin of non-monotonicity in mean-squared displacement}
\label{sec:localized_bright_state}
\begin{figure}[!htpb]
\centering
\includegraphics[width=\linewidth]{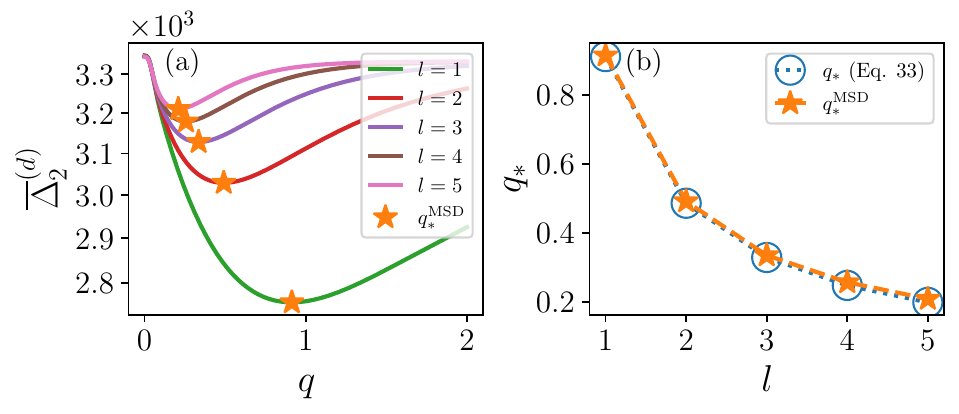}
\caption{(a) Infinite-time-averaged MSD as a function of defect strength $q$ for different distances $l$ between the initial and defect sites. Orange stars mark the numerical minima $q_*^{\rm MSD}$. (b) Overlap-optimal strength $q_*$ from Eq.~\eqref{eq:q*} (blue circles) and the numerical MSD minima $q_*^{\rm MSD}$ (orange stars) as functions of $l$. The infinite-time average is calculated from the spectral decomposition using eigenvectors obtained by exact diagonalization~\cite{data}. Both panels use $N=200$, $\gamma=1$, and $\ket{n_0}=\ket{2}$. The lines in (b) are guides for the eye.}
\label{fig:Fig2}
\end{figure}

In this section, we uncover the origin of the non-monotonic behavior of the MSD from the opposing contributions of the bright localized state and the bright extended states. Figure~\ref{fig:Fig2}(a) displays non-monotonic behavior of MSD as a function of defect strength $q$ for several distances between the initial and defect sites (See also Ref.~\cite{AcharyaGiuggioliGupta}). The goal of this section is to provide an estimate of the characteristic defect strength at which the MSD attains a minimum, as shown by the orange stars in Fig.~\ref{fig:Fig2}(a).
We will start by looking at the contribution of the bright localized state, which satisfies $E_{\rm loc}<-2\gamma$. In the thermodynamic limit ($N\to\infty$), we have
\begin{align}
    F(E_{\rm loc})
    &=\frac{1}{2\pi}\int_0^{2\pi}
    \frac{dk}{-E_{\rm loc}-2\gamma\cos k}
    =\frac{1}{\sqrt{E_{\rm loc}^2-4\gamma^2}}.
\end{align}
The self-consistency condition $qF(E_{\rm loc})=1$ therefore gives
$E_{\rm loc}=-\sqrt{4\gamma^2+q^2}$, which implies that there is only one localized bright state. For convenience, we write $q=2\gamma\sinh\kappa$, which yields $E_{\rm loc}=-2\gamma\cosh\kappa$ with $\kappa>0$. The components of the localized eigenvector with eigenvalue $E_{\rm loc}$ satisfy
\begin{align}
-\gamma\bigl(u_{n+1}(E_{\rm loc})+u_{n-1}(E_{\rm loc})\bigr)
    &=E_{\rm loc}\,u_n(E_{\rm loc}),\\
-\gamma\bigl(u_{n_d+1}(E_{\rm loc})+u_{n_d-1}(E_{\rm loc})\bigr)
    -qu_{n_d}(E_{\rm loc})
    &=E_{\rm loc}\,u_{n_d}(E_{\rm loc})
\end{align}
away from and at the defect site, respectively. Since $E_{\rm loc}<-2\gamma$, the localized eigenstate takes the thermodynamic-limit form
$u_n(E_{\rm loc})=\mathcal{N}e^{-\kappa d(n,n_d)}$, which peaks at the defect site and decays exponentially with distance from it (see Fig.~\ref{fig:Fig1}(b)). Here $d(n,n_d)\equiv\min\{|n-n_d|,N-|n-n_d|\}$. At the defect site, this form gives $q=2\gamma\sinh\kappa$. Thus, the normalization condition $\sum_n|u_n(E_{\rm loc})|^2=1$ gives
\begin{align}
1
&=\mathcal{N}^2\left(1+2\sum_{d=1}^{\infty}e^{-2\kappa d}\right)
 =\mathcal{N}^2\left(1+2\frac{e^{-2\kappa}}{1-e^{-2\kappa}}\right)
 =\mathcal{N}^2\coth\kappa,
\end{align}
and hence
\begin{align}
    u_n(E_{\rm loc})
    =\sqrt{\tanh\kappa}\,e^{-\kappa d(n,n_d)}.
    \label{eq:localized_eigenvector}
\end{align}
Note that on a finite ring, the exact localized profile contains contributions from both paths around the ring:
$u_n^{(N)}\propto e^{-\kappa_N d(n,n_d)}
+e^{-\kappa_N[N-d(n,n_d)]}$, where
$q=2\gamma\sinh\kappa_N\tanh(N\kappa_N/2)$.
Equation~\eqref{eq:localized_eigenvector} follows when $N\kappa_N\gg1$.

For a particle initially localized at $n_0$, its overlap with the localized state gives
\begin{align}
    O_{\rm loc}(l,q)
    &=|\langle E_{\rm loc}|n_0\rangle|^2 = \vert u_{n_0}(E_{\rm loc})\vert^2=\tanh\kappa\,e^{-2\kappa l},
    \label{eq:bound_state_overlap}
\end{align}
where $l=d(n_0,n_d)$. When $l=0$, this weight increases monotonically with $q$. For $l>0$, it is non-monotonic and reaches its maximum at $\sinh(2\kappa_*)=1/l$, giving the overlap-optimal defect strength
\begin{align}
    q_*
    =2\gamma\sinh\left[
        \frac{1}{2}\operatorname{arcsinh}\left(\frac{1}{l}\right)
      \right].
    \label{eq:q*}
\end{align}
Thus, when the particle starts away from the defect, its overlap with the localized state is largest at $q=q_*$. Since the dark state overlap with the initial site is independent of $q$, the $q$-dependence comes solely from the bright state overlap with the initial site. For $n_d \neq n_0$ and at $q = q_*$, the overlap with the initial site is maximal for a bright localized state which tends to localize the particle and is therefore minimal for the total bright extended state subspace which spreads the particle all over the lattice. This suggests that the mean square displacement is expected to be minimum at this particular defect strength $q = q_*$.  Numerically, we found excellent agreement between the minimum of the time-averaged MSD with the analytically estimated overlap-optimal $q_*$ in Eq.~\eqref{eq:q*} for different distances between the initial and defect site for finite $N$ as shown in Fig.~\ref{fig:Fig2}(b). 

\begin{figure}[!htpb]
\centering
\includegraphics[width=\linewidth]{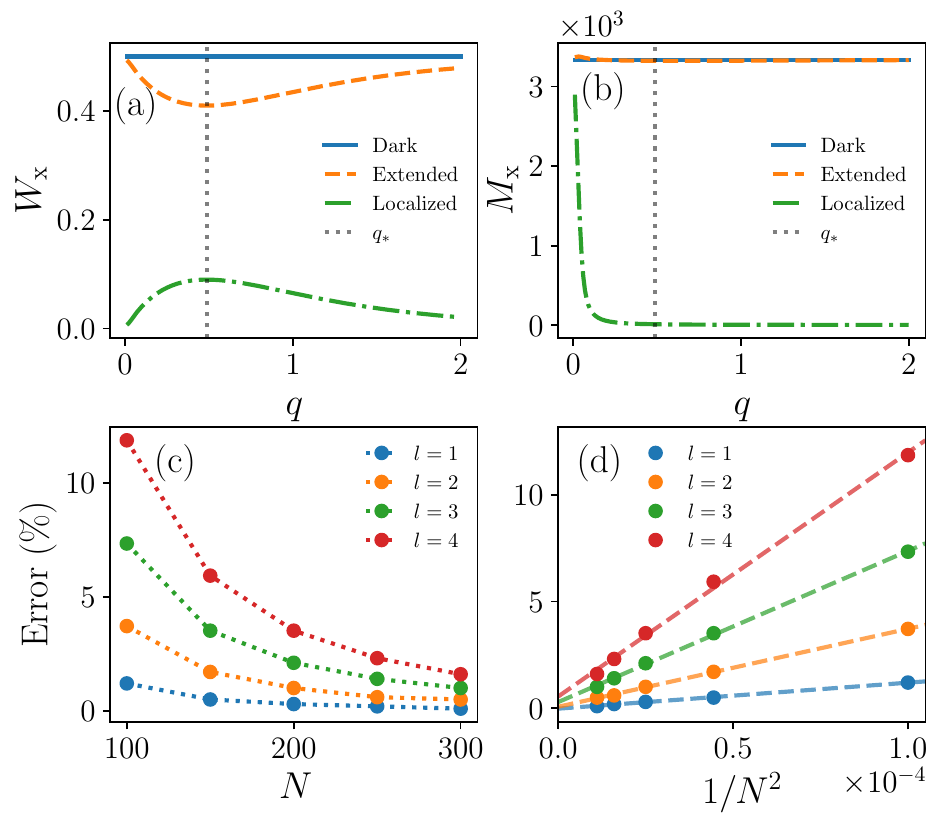}
\caption{(a) Initial-state weights $W_x$ and (b) conditional second moments $M_x$ of the dark, bright extended, and bright localized sectors as functions of $q$. The vertical dotted line indicates the overlap-optimal strength $q_*$ from Eq.~\eqref{eq:q*}. Panels (a) and (b) use $N=200$, $\gamma=1$, $\ket{n_0} = \ket{2}$, and $l=2$. (c) Percent relative error as a function of $N$ for $l=1,\ldots,4$. (d) Percent relative error plotted against $1/N^2$. Other numerical parameters are $\gamma = 1$ and $\ket{n_0} = \ket{2}$. The dashed lines are fitted using NumPy polyfit function~\cite{data}.}
\label{fig:Fig3}
\end{figure}

To quantify how this redistribution of overlap weight affects the MSD, we also need the spatial second moment of each sector. Let $x\in\{\mathrm{D},\mathrm{ext},\mathrm{loc}\}$ label the dark, bright extended, and bright localized sectors, respectively. For each sector, we define its initial-state weight $W_x$ and conditional second moment $M_x$ by
\begin{align}
W_x(q)
    &=\sum_{j\in\mathcal{H}_x}
      |\langle\phi_j|n_0\rangle|^2, \nonumber\\
M_x(q)
    &=\frac{1}{W_x(q)}
      \sum_{j\in\mathcal{H}_x}
      |\langle\phi_j|n_0\rangle|^2
      \sum_{n=0}^{N-1}d(n,n_0)^2|\langle n|\phi_j\rangle|^2,
      \nonumber\\
\overline{\Delta}_2^{(d)}
    &=\sum_x W_xM_x.
    \label{eq:sector_weights_moments}
\end{align}
For the localized sector, the sum contains only $\ket{E_{\rm loc}}$. Thus $W_{\rm loc}=O_{\rm loc}$ in the thermodynamic limit, while $W_{\rm D}$ is independent of $q$ and $W_{\rm ext}=1-W_{\rm D}-W_{\rm loc}$. Figure~\ref{fig:Fig3}(a) shows that the localized-state weight reaches a maximum near $q_*$, while the extended-sector weight reaches a minimum and the dark state weight remains independent of $q$.
Although the bound state is localized near the defect, its second moment about the initial site is not zero. When $l\ll N$ and $N\kappa\gg1$, the ring can be treated as an infinite chain. Choosing $n_d=0$ and $n_0=l$, Eq.~\eqref{eq:localized_eigenvector} gives
\begin{align}
M_{\rm loc}(l,q)
&=\tanh\kappa\sum_{n=-\infty}^{\infty}
(n-l)^2e^{-2\kappa|n|} \nonumber\\
&=l^2+2\tanh\kappa\sum_{n=1}^{\infty}
n^2e^{-2\kappa n} =l^2+\frac{1}{2\sinh^2\kappa}
=l^2+\frac{2\gamma^2}{q^2}.
\label{eq:localized_second_moment}
\end{align}
The term proportional to $n$ vanishes because the bound-state probability is symmetric about the defect. The $l^2$ term accounts for the separation between the defect and initial sites, while the remaining term measures the finite width of the bound state. $M_{\rm loc}$ is much smaller than the extended-sector moment $M_{\rm ext}\sim N^2$. In Fig.~\ref{fig:Fig3}(b), we plot $M_x$ as a function of $q$. $M_{\rm loc}$ is large only at small $q$ and then monotonically decreases to the value of $l^2$ with increasing $q$ as given by Eq.~\eqref{eq:localized_second_moment}. Note also that the second moments change comparatively little close to $q_*$. Since the dark-sector weight is independent of $q$ and $W_{\rm ext}=1-W_D-W_{\rm loc}$, MSD can be rewritten as
\begin{align}
    \overline{\Delta}_2^{(d)}(q)
= W_{\rm D} M_{\rm D} + (1 - W_{\rm D})M_{\rm ext}(q) +\bigl[M_{\rm loc}(q)-M_{\rm ext}(q)\bigr]W_{\rm loc}(q)
\label{eq:MSD_approx}
\end{align}
Since the sector moments vary comparatively little near $q_*$, and $M_{\rm loc}(q)<M_{\rm ext}(q)$, increasing the localized-state weight thus lowers the MSD. Its maximum therefore predicts a minimum near $q_*$.

Note that the estimate of $q_*$ is derived in the thermodynamic limit. For finite $N$, there is a small offset between $q_*$ and the numerical MSD minimum as shown in Fig.~\ref{fig:Fig2}(b). We quantify the offset using the percent relative error, defined by $100 \times (\vert q_*^{\rm MSD}-q_*\vert)/q_*$. Figure~\ref{fig:Fig3}(c) shows how the percent relative error depends on $l$ and $N$. For each fixed $l$, this relative error decreases as $N$ increases. In Fig.~\ref{fig:Fig3}(d), we show that the error decreases approximately like $N^{-2}$. This scaling can be intuitively understood as follows.  From Eq.~\eqref{eq:MSD_approx}, one can infer that at $q_*$,  
\begin{align}
    \frac{\mathrm{d}\overline{\Delta}_2^{(d)}(q)}{\rm{d} q}\Bigg\vert_{q = q_*} = (1 - W_{\rm D})M^{\prime}_{\rm ext}(q_*)+ W_{\rm loc}(q_*) [M^\prime_{\rm loc}(q_*) - M^\prime_{\rm ext}(q_*)] \approx \mathcal{O}(1)
\end{align}
where by definition, the first-order variation of the localized-state weight vanishes i.e. $W^\prime_{\rm loc}(q_*) = 0$. For fixed $l$, Eq.~\eqref{eq:q*} shows that $q_*$ is independent of $N$. We further assume, consistently with the weak variation of sector moments observed in Fig.~3(b), that their derivatives do not scale with $N$. Now, let's assume the actual minima of MSD occurs at $q_* + \delta q$ for finite $N$, where $\delta_q$ is the small offset. Now Taylor expanding upto first order in $\delta q$ gives
\begin{align}
    &\frac{\mathrm{d}\overline{\Delta}_2^{(d)}(q)}{\rm{d} q}\Bigg\vert_{q = q_* + \delta q} \approx \frac{\mathrm{d}\overline{\Delta}_2^{(d)}(q)}{\rm{d} q}\Bigg\vert_{q = q_* } +  \frac{\mathrm{d}^2\overline{\Delta}_2^{(d)}(q^)}{\rm{d} q^2}\Bigg\vert_{q = q_* }\, \delta q = 0 
    \label{eq:derivative_condition}
\end{align}
and the double derivative follows
\begin{align}
     \frac{\mathrm{d}^2\overline{\Delta}_2^{(d)}(q)}{\rm{d} q^2}\Bigg\vert_{q = q_* }  &=   (1 - W_{\rm D})M^{\prime\prime}_{\rm ext}(q_*) +  W_{\rm loc}(q_*) [M^{\prime \prime}_{\rm loc}(q_*) - M^{\prime \prime}_{\rm ext}(q_*)] \nonumber \\ &+  W^{\prime \prime}_{\rm loc}(q_*) [M_{\rm loc}(q_*) - M_{\rm ext}(q_*)] 
\end{align}
The third term is of $\mathcal{O}(N^2)$ because  $W_{\rm loc}^{\prime \prime}(q_*)=\mathcal{O}(1)$, $M_{\rm loc}(q_*)=\mathcal{O}(1)$, and $M_{\rm ext}(q_*)$ scales as $\mathcal{O}(N^2)$. The first and second terms are subleading and therefore does not alter the leading $\mathcal{O}(N^2)$ scaling of the curvature. Equation~\eqref{eq:derivative_condition} thus balances an $\mathcal{O}(1)$  slope against an $\mathcal{O}(N^2)\delta q$ correction, which requires $\delta q=\mathcal{O}(N^{-2})$.

\section{Conclusions}
We provide a complementary spectral derivation of the dynamics of the nearest-neighbour tight-binding model with a single defect~\cite{AcharyaGiuggioliGupta}. The eigenstate formulation makes the physical structure of the dynamics particularly transparent. The Hilbert space decomposes naturally into a dark subspace, whose states have zero amplitude at the defect site and are therefore unaffected by it, and a bright subspace, which couples to the defect and contains the entire defect-induced dynamics. The spectral formulation not only reproduces the exact dynamics but also reveals that the minimum of the MSD occurs close to the defect strength for which the overlap of the initial state with the bright localized state is maximal, thereby explaining the non-monotonic dependence of the MSD on the defect strength. The formalism can be naturally extended to the case of multiple defect sites, as discussed in \ref{app:6}. The spectral approach may be particularly useful for problems in which state-resolved information is required. For example, dark-state polaritons, which are analogous to the dark states identified here, enable quantum memory protocols~\cite{FleischhauerLukin}.
\section*{Acknowledgements}
    The author acknowledges discussions with and helpful comments of Anish Acharya, Shamik Gupta and Giovanna Morigi. The author also acknowledges Sagnik Chatterjee for help with figures. This work was funded by the German Ministry of Education and Research (BMBF, Project ``NiQ: Noise in Quantum Algorithms"), by the Deutsche Forschungsgemeinschaft (DFG, German Research Foundation) – Project-ID 429529648 – TRR 306 QuCoLiMa (``Quantum Cooperativity of Light and Matter''), and by the QuantERA II Programme (project ``QNet: Quantum transport, metastability, and neuromorphic applications in Quantum Networks"), which has received funding from the EU's Horizon 2020 research and innovation programme under Grant Agreement No.\ 101017733, as well as from the Deutsche Forschungsgemeinschaft DFG (Project ID 532771420).

\section*{Data Availability}
The data and code used in this work are available in the v2.0.0 release of the GitHub repository at \url{https://github.com/sayanr5834/NN_single_defect} and archived at~\cite{data}.
\section*{References}
\bibliographystyle{iopart-num}
\bibliography{reference}

\appendix
\section{Common factors in Eq.~\eqref{eq:secular_chebyshev}}
\label{app:1}
For even $N = 2m$, using the product identities of Chebyshev polynomials~\cite{NIST:DLMF}, one can write
\begin{align}
    T_{2m}(x) + 1 &= 2T_m^2(x) - 1+ 1 = 2T_m^2(x),\\
    U_{2m - 1}(x) &= 2T_m(x)U_{m-1}(x),
\end{align}
which gives Eq.~\eqref{eq:secular_chebyshev} as
\begin{align}
    1 - \frac{q}{2 \gamma}\frac{T_m(x)}{(x^2 - 1)U_{m-1}(x)} = 0\,.
\end{align}
This gives the polynomial expression $Q_{B}^{\text{even}}(x) = (x^2 - 1)U_{m-1}(x) - (q/2\gamma)T_{m}(x)$ for even $N$, whose roots correspond to $m +1$ bright eigenvalues.

For odd $N = 2m + 1$, one finds~\cite{NIST:DLMF}
\begin{align}
     T_{2m+ 1}(x) + 1 &= (x + 1) (U_m(x) - U_{m-1}(x))^2, \\
     U_{2m}(x) &= U_m^2(x) - U^2_{m-1}(x),
\end{align}
which gives Eq.~\eqref{eq:secular_chebyshev} as
\begin{align}
    1 -  \frac{q}{2\gamma} \frac{U_m(x) -U_{m-1}(x)}{(x - 1)(U_{m}(x) + U_{m-1}(x))} = 0\,.
\end{align}
This gives the polynomial expression $Q_{B}^{\text{odd}}(x) = (x - 1)(U_m(x) + U_{m-1}(x)) - (q/2\gamma)(U_m(x) - U_{m-1}(x))$ for odd $N$, whose roots correspond to $m +1$ bright eigenvalues.
\section{Recasting $f_j$ in terms of Chebyshev polynomials}
\label{app:2}
We show that $f_j = P(x_j)/Q^{\prime}(x_j)$ as in \cite{AcharyaGiuggioliGupta}, where $P(x)\equiv T_{N-|n_0 - n_d|}(x)+T_{|n_0 - n_d|}(x)$ and $Q(x)$ is given by Eq.~\eqref{eq:eigval}. Using Eq.~\eqref{eq:bright}, one finds
\begin{align}
    f_j  &= \frac{1}{q F^\prime(E_j)} \frac{1}{N} \sum_{r = 0}^{N - 1} \frac{e^{-i k_r (n_0 - n_d)}}{\epsilon_{k_r} - E_j} = \frac{1}{q F^\prime(E_j)} \frac{1}{2\gamma N}
\sum_{r=0}^{N-1}\frac{e^{-ik_rl}}{x_j-\cos k_r} \nonumber \\
&= \frac{1}{q F^\prime(E_j)} \frac{1}{2\gamma N} \sum_{r=0}^{N-1}\frac{\cos(k_r \,l)}{x_j-\cos k_r}\,,
\end{align}
where we have used the fact that the imaginary part vanishes because of the contribution of $k_r$ with $k_{N-r}$ and denoted $l = d(n_0,n_d)$. One finds $T_{N-l}(\cos k_r)+T_l(\cos k_r) =\cos[(N-l)k_r]+\cos(l k_r) = 2\cos(l k_r)$. Hence one obtains 
\begin{align}
\frac{1}{N}
\sum_{r=0}^{N-1}
\frac{\cos(lk_r)}{x_j-\cos k_r}
=
\frac{T_{N-l}(x_j)+T_l(x_j)}
{(x_j^2-1)U_{N-1}(x_j)}.
\label{eq:cheb_sum_l}
\end{align}
Note that for the case $l = 0$, the numerator becomes $T_N(x) + T_0(x) = T_N(x) +1$. The coefficient $f_j$ can thus be rewritten as
\begin{align}
f_j &=\frac{1}{qF'(E_j)}\frac{1}{2 \gamma}\frac{T_{N-l}(x_j)+T_l(x_j)}{(x_j^2-1)U_{N-1}(x_j)} = \frac{1}{qF'(E_j)}\frac{P(x_j)}{2 \gamma D(x_j)},
\label{eq:fj}
\end{align}
where we define $D(x)\equiv (x^2-1)U_{N-1}(x)$. Using Eqs.~\eqref{eq:secular_chebyshev} and ~\eqref{eq:eigval}, we obtain
\begin{align}
F(E)&=\frac{T_N(x)+1}{2\gamma D(x)}, \\
Q(x)&= D(x)-\frac{q}{2\gamma}[T_N(x)+1],
\end{align}
and therefore $qF(E)=1- Q(x)/D(x)$. Since $x=-E/(2\gamma)$, at a bright eigenvalue $E_j=-2\gamma x_j$ satisfying $Q(x_j)=0$, one obtains
\begin{align}
qF'(E_j)=\frac{Q'(x_j)}{2\gamma D(x_j)}.
\label{eq:Fprime}
\end{align}
Note that here, $F^\prime(E)$ denotes differentiation with respect to $E$, whereas $Q^\prime(x)$ denotes differentiation with respect to $x$. Substituting Eq.~\eqref{eq:Fprime} into Eq.~\eqref{eq:fj} thus gives $f_j= P(x_j)/Q'(x_j)$. Inserting this expression into the definitions in Sec.~\ref{sec:eigvec} reproduces the form of $I_{n}^{(d)}(t)$ and $K_{n}^{(d)}(t)$ as  given in Eq.~(C.1) and Eq.~(C.2) of Ref.~\cite{AcharyaGiuggioliGupta}.


\section{The case of multiple defects}
\label{app:6}
The secular-equation approach can be extended to the case of multiple defects. Consider $M$ defects located at sites $d_a$, with $a=1,\ldots,M$, and defect strengths $q_a>0$. The Hamiltonian is
\begin{align}
\hat H=\hat H_0-\sum_{a=1}^{M} q_a\ket{d_a}\bra{d_a}.
\end{align}
Let $\ket E$ be an eigenstate of $\hat H$, and define its amplitudes on the defect sites as
$u_a(E)\equiv \langle d_a \vert E \rangle$. For an eigenstate with at least one nonzero defect-site amplitude $u_a(E)$, one finds the bright eigenstate as
\begin{align}
\ket E = \sum_{a=1}^{M}\frac{q_a u_a(E)}{\hat H_0-E}\ket{d_a}.
\end{align}
Projecting this equation onto each of the defect sites gives a closed set of $M$ linear equations. Defining $F_{ba}(E) \equiv \bra{d_b}(\hat H_0-E)^{-1}\ket{d_a}$, one can now see that the bright-state eigenvalues are given by the self-consistency condition
\begin{align}
\det \left[\delta_{ba}-q_a F_{ba}(E)\right]=0.
\end{align}
\end{document}